\documentclass[a4paper]{jpconf}
\usepackage{graphicx}

\def\be{\begin{equation}}
\def\ee{\end{equation}}
\def\bea{\begin{eqnarray}}
\def\eea{\end{eqnarray}}
\def\ssA{{\scriptscriptstyle A}}
\def\ssM{{\scriptscriptstyle M}}
\def\ssN{{\scriptscriptstyle N}}
\def\ssB{{\scriptscriptstyle B}}
\def\ssP{{\scriptscriptstyle P}}
\def\ssQ{{\scriptscriptstyle Q}}
\def\ssR{{\scriptscriptstyle R}}
\def\ssI{{\scriptscriptstyle I}}
\def\ssS{{\scriptscriptstyle S}}
\def\ssKK{{\scriptscriptstyle KK}}
\def\ssJ{{\scriptscriptstyle J}}
\def\Dsl{\,\raise.15ex\hbox{/}\mkern-13.5mu D}

\begin{document}

\hspace{8cm} FTUAM-12-104, IFT-UAM/CSIC-12-92

\title{Bulk Renormalization and Particle Spectrum in Codimension-Two Brane Worlds}

\author{Alberto Salvio}

\address{Scuola Normale Superiore and INFN, Piazza dei Cavalieri 7, 56126 Pisa, Italy

\vspace{0.2cm}
Departamento de F\'isica Te\'orica,  Universidad Aut\'onoma de Madrid and \\ 
Instituto de F\'isica Te\'orica IFT-UAM/CSIC, Cantoblanco, 28049 Madrid, Spain

 }

\ead{alberto.salvio@sns.it}

\begin{abstract}   
    We study the Casimir energy due to bulk loops of matter fields in codimension-two brane worlds and 
    discuss how effective field theory methods allow us to use this result 
    to renormalize the bulk and brane operators.  In the calculation we explicitly sum over the Kaluza-Klein (KK) states 
    with a new convenient method, which is based on a combined use of zeta function and dimensional regularization.
      Among the general class of models we consider we include a supersymmetric example,
   6D gauged chiral supergravity. 
    Although much of our discussion is more general, we treat in some detail a class of compactifications,
   where the extra dimensions parametrize a rugby ball shaped space with size stabilized by  a bulk magnetic  flux.       The rugby ball geometry requires two branes, which can host the Standard Model fields and carry both tension and magnetic flux (of the bulk gauge field), the leading terms in a derivative expansion. The brane properties have an impact on the KK spectrum and therefore on the Casimir energy as well as on the renormalization of the brane operators.   A very interesting feature is that when the two branes carry exactly the same amount of flux, one half of the bulk supersymmetries survives after the compactification, even if the brane tensions are large.
      We also discuss the implications of these calculations for the natural value of the cosmological constant  when the bulk has two large extra dimensions and the bulk supersymmetry is partially preserved  (or completely broken).

\end{abstract}

\section{Introduction}

Higher dimensional theories have a variety of motivations in the physics of fundamental interactions. The original one, in the pioneering works of Kaluza and Klein (KK), was the unification of the known forces. This idea survived as the decades passed by through superstring theory, which provides a framework where gravity can be unified with the  other interactions and strongly indicates that the space-time contains a number of extra dimensions. Other advantages of higher dimensional setups include the possibility to understand the structure of fermion masses and couplings \cite{Grossman:1999ra} in the Standard Model. 

Remarkably, extra dimensions also provide mechanisms to understand technical naturalness problems. For example, 
the gauge hierarchy problem, which queries why the Fermi scale is so much smaller than the Planck mass, can be addressed in the large extra dimension (LED) scenario
or by so called warped geometries.  It is then natural to ask whether extra dimensions can help us understand the
most serious fine tuning issue we know,  the cosmological constant problem \cite{Weinberg:1988cp}.

To understand the origin of such problem it is useful to look at the usual Einstein-Hilbert 
gravity\footnote{We adopt the mostly plus signature and  the  curvature conventions of Ref. \cite{G&C}.}
coupled to matter 
\begin{equation} 
 \frac{{\cal L}}{\sqrt{-g}}= -\frac{R}{2 \kappa_4^2} - \Lambda_0 + \frac{{\cal L}_{matter}}{\sqrt{-g}}\, ,
\end{equation}%
where ${\cal L}_{matter}$ is the part which depends on the matter fields.
This represents the leading gravitational tensor theory in the large wavelength limit. Here $\kappa_4$ is the 4D Planck scale
and $\Lambda_0$ the tree level value of the cosmological constant. From this equation it is clear that the 
quantum vacuum energy $\langle \rho \rangle$ produced by the gravitational field $g_{\mu \nu}$ and any matter particle contributes to the cosmological 
constant $\Lambda= \Lambda_0+\langle \rho \rangle$. The cosmological constant problem consists in the mismatch between
the observed value $\Lambda \sim (10^{-3}\mbox{eV})^4$ and the (relatively) enormous contribution to $\langle \rho \rangle$ due to 
the known forms of matter: any Standard Model particle with mass $m$ adds to $\Lambda$ a quantity  of  order $m^4$ and therefore a huge fine tuning of $\Lambda_0$ is required to obtain the observed value.

Any solution of such problem must involve a modification of gravity at 
an energy scale roughly of order $10^{-3}$eV. Indeed we know all non-gravitational forces up to the $\sim$ 10 TeV scale
and the only possible modification should therefore occur in the gravity sector, which is much less constrained by the 
experiments; if such modification emerges only at some intermediate energy scale $E_{int}$, between $10^{-3}$eV and 10 TeV,
then the low energy effective theory at energies much smaller than $E_{int}$ would be again the Standard Model 
plus Einstein's gravitational theory and we would be back to the original problem.

The required modification occurs in any model with two large extra dimensions. 
Let us see why. The LED scenario predicts the following ratio between the 4D Planck mass, $M_{\rm Pl}\equiv 1/(\sqrt{2} \kappa_4)$, and its higher dimensional
counterpart, $M\equiv 1/(\sqrt2 \kappa)^{2/(D-2)}$, \cite{LED}
\be \frac{M_{\rm Pl}^2}{M^2}=(Mr)^{D-4}\, ,\ee
where $D$ is the full space-time dimension and $r$ is the linear size of the $D-4$ dimensional space volume (that is the length scale of the extra dimensions).
In this framework one addresses the gauge hierarchy problem by choosing roughly $M\sim$ TeV, which for $D=6$ gives us a KK scale
of order $1/r \sim 10^{-3}$eV (above which gravity is modified). It is important to notice that a true solution
of the gauge hierarchy problem requires a mechanism to dynamically stabilize $r$ to this large value;
one way to do so is to consider a compactification on a topologically non-trivial space with a 
magnetic flux (flux stabilization). Also, notice that in order for the extra dimensions to be so large,  only gravity (among the known interactions) can propagate in the bulk 
while the Standard Model fields should be confined on a 3+1 dimensional brane (3-brane henceforth), which in
the case of interest here, $D=6$, has to be of  codimension-two.   

Of course, the fact that gravity is modified at the required energy scale is not by itself a solution of the cosmological
constant problem. We need a further mechanism protecting $\Lambda$ from large quantum corrections. Supersymmetry is a 
possible candidate because the fermion and boson contributions to the  vacuum energy cancel exactly if supersymmetry 
is unbroken, or, if supersymmetry is broken at an energy scale $m_\ssS$, produce a net vacuum energy of order $m_\ssS^4$. A crucial
observation now is that the supersymmetry breaking scale in the bulk does not have to be of the same order of that 
on the brane, which is required to be larger than the TeV scale by the LHC experiments. A supersymmetry breaking brane
Lagrangian $\delta  L_b$, which, including quantum corrections, is expected to be of order TeV gives rise to a 
supersymmetry breaking  splitting $\delta m_{\ssKK}^2$ in the KK spectrum of order $\kappa^2 \delta  L_b/r^2$. Here the 
factor $\kappa^2$ reminds us that the brane physics is communicated to the  bulk through gravitational interactions
while  $1/r^2$ is there for dimensional reasons. To the extent that $m_\ssS^2$ is given by $\delta m_{\ssKK}^2$
we therefore obtain a cosmological constant of the correct order of magnitude. To summarize
$$ m_\ssS^2 \stackrel{\scriptscriptstyle{?}}{\sim} \delta m_{\ssKK}^2 \sim  
\frac{\kappa^2 \delta L_b}{r^2} \stackrel{\scriptscriptstyle{LED}}{\sim}
 \frac{1}{r^2}\,,$$
where the question mark reminds us our expectation. To confirm it explicit calculations in a concrete model are therefore 
needed. 

The idea that supersymmetric large extra dimensions with codimension-two branes can address the gauge hierarchy
and the cosmological constant problem was originally proposed in \cite{Aghababaie:2003wz}. We refer to these 
works for an extended discussion. But the fact that extra dimensions in general 
can help with the cosmological constant problem is 
older (see for example Refs. \cite{Rubakov:1983bz})  

Since any issue of technical naturalness is a quantum problem we believe it is important to have a sistematic approach to compute quantum corrections in codimension-two brane worlds.
Here we discuss quantum corrections due to bulk (massive) states, focusing on the one-loop approximation. The main subject of this article is the Casimir energy produced by integrating out a (massive) bulk field, and how to obtain it from its KK mass spectrum. Much of our discussions  is based on Refs. \cite{Parameswaran:2006db,Parameswaran:2007cb,Parameswaran:2009bt,articleA,articleB}, but here we also obtain some original results.  The main one is a convenient way to compute Casimir energies from KK spectra, which is based on a combined use of dimensional regularization (for the ultraviolet divergences) and zeta function regularization (for the KK sums).    The Casimir energy can then be used to calculate the renormalization group equations (RGEs) of both bulk and brane coefficients in the quantum action.

Although some results can be applied to any codimension-two model, we study in some detail a concrete class of compactifications, where the extra dimensions form a rugby ball shaped space and the flux of a gauge field stabilizes their size, $r$. These solutions are supported by two 3-branes having  both tension \cite{conical} and magnetic flux (the leading terms in a derivative expansion of the brane action) \cite{localizedflux}. We discuss a class of models having these solutions, including one with bulk supersymmetry, 6D gauged chiral supergravity \cite{MS,NS,SS}, which we refer to throughout this paper as our supersymmetric example.
As shown in Ref. \cite{articleB}, when the two 3-branes have identical localized fluxes one half of the bulk supersymetries are preserved (if the bulk is supersymmetric), implying, interestingly, that the 4D vacuum energy vanishes in this limit.
 
 For  rugby ball compactifications, the KK towers of many bulk fields are known from previous calculations  \cite{Parameswaran:2006db,Parameswaran:2007cb,Parameswaran:2009bt,articleA,articleB} and we make use of these results to compute explicitly the Casimir energy and the renormalization of bulk and brane coefficients as a function of the bulk mass $m$ of the field we integrate out.
 Our computational method  is very convenient and confirms the results of Ref. \cite{articleA}.  
 The final expression for the Casimir energy is a polynomial function  in $m r$ of degree 6, 
 where the coefficients depend on the brane tension and the bulk and brane localized fluxes. In the supersymmetric
 case, all these coefficients vanish for identical localized fluxes or, more generically, they are suppressed by the 
 difference between these fluxes. As shown in \cite{articleB} the 4D cosmological constant inherits this suppression and
 can  be of the observed size.
 
Let us give the outline of this article. In section \ref{class} we introduce our family of models, including both
the bulk and brane actions, and the rugby ball solutions; as a concrete supersymmetric example we define 6D gauged
chiral supergravity and reexamine the technical naturalness of the cosmological constant in this specific case. We then
review the KK spectra of scalars, fermions and vectors for rugby ball compactifications in section  \ref{spectra}.
The Casimir energy calculation and the bulk renormalization is performed in section \ref{Casimir-renormalization},
treating in some detail the codimension-two case and, more specifically, the rugby ball solutions. 
In section \ref{Casimir-renormalization-RB} we finally discuss the form of the Casimir energy due to loops of specific matter fields: the simplest case of a  
real scalar and (massive) matter multiplets of 6D supergravity.

\section{The class of models} \label{class}

We focus on a class of models which include, in addition to the metric tensor $g_{\ssM\ssN}$, a set of gauge fields
$A^a_\ssM$, scalars $\phi^i$ and fermions $\psi^r$. The bosonic part of the Lagrangian for these fields
is
\be
\label{E:Baction}
    \frac{{\cal L}_\ssB}{\sqrt{- g}} = -\, \frac{1}{2\kappa^2}  R - \frac{1}{2} \, {\cal G}_{ij}(\phi)
    D_{\ssM} \phi^i \, D^\ssM \phi^j  - \frac{1}{4} {\cal H}_{ab}(\phi) \; F^a_{\ssM\ssN} F^{\ssM\ssN b} -  V(\phi) \,,
\ee
where $D_\ssM$ is the gauge-covariant derivatives for the scalars, 
$F_{\ssM\ssN}^a$ is the field strength of $A^a_\ssM$, 
and ${\cal G}_{ij}(\phi)$, ${\cal H}_{ab}(\phi)$ and $V(\phi)$ are generic functions of the scalars.

This is general enough to describe the linearized dynamics of matter supermultiplets in
explicit 6D supergravities. The supersymmetric example we shall refer to throughout this paper is 6D gauged chiral 
supergravity. However, the general class of models we consider also includes non-supersymmetric theories and 
the results we shall discuss in the following sections  hold for them as well (unless otherwise stated).
The supersymmetric field content is given by a
supergravity-tensor multiplet ($g_{\ssM\ssN}$, $B_{\ssM\ssN}$, $\phi$,
$\psi_\ssM$, $\chi$)  - with metric tensor, anti-symmetric Kalb-Ramond field
(with field strength $G_{\ssM\ssN\ssP}$), dilaton, gravitino and dilatino -
coupled to some gauge multiplets ($A^a_\ssM$, $\lambda$) - with
gauge fields and gauginos - and some hypermultiplets
($\Phi^I$, $\Psi$) - each of them having hyperscalars and their chiral hyperini.  The fermions
are all Weyl spinors and satisfy $\Gamma_7 \psi_\ssM = \psi_\ssM$,
$\Gamma_7 \chi = - \chi$, $\Gamma_7 \lambda = \lambda$ and
$\Gamma_7 \Psi = - \Psi$.  We will consider a
matter content with gauge group of the form
$\mathcal{G}=\tilde{\mathcal{G}}\times U(1)_R$, where $U(1)_R$ is an Abelian R-symmetry,
and $\tilde{\mathcal{G}}$ is a generic product of simple groups.
The bosonic Lagrangian of 6D gauged chiral supergravity is given by
\be
    \frac{{\cal L}_\ssB}{\sqrt{- g}}
    = -\, \frac{1}{2\kappa^2} ( R +
     \partial_{\ssM} \phi \, \partial^\ssM \phi)
    - \frac{e^{-\phi}}{4g_a^2} \; F^a_{\ssM\ssN} F_a^{\ssM\ssN}- \frac12 \, G_{\ssI \ssJ}(\Phi) \, g^{\ssM \ssN} D_\ssM \Phi^\ssI D_\ssN \Phi^\ssJ - \frac{2 g^2}{\kappa^4} \,  e^\phi \, U(\Phi) \,, \label{Lagrangian-class}
\ee
where $g$ is the $U(1)_R$ gauge coupling and we have set $G_{\ssM\ssN\ssP}=0$ for simplicity.

As we have already mentioned a realistic realization of the large extra dimensions idea requires the 
presence of 3-branes, one of which supports the Standard Model fields. We take the brane action
to be \cite{localizedflux}.
 \bea \label{eq:genbraneaction}
 S_b &=& - \int d^4x \, \sqrt{-g} \; L_b \nonumber\\
\hbox{with} \quad L_b &=& T_b - \frac{ \mathcal{A}_b}{2 g_b^2} \, 
\epsilon^{mn} F_{mn} + \frac{\mathcal{B}_b}{\kappa} \, R + \frac{\mathcal{C}_b}{\kappa}
(\partial \phi)^2 + \cdots \,,
 \eea
where $g_b$  is the gauge coupling of the gauge field corresponding to the second term in (\ref{eq:genbraneaction})
(which has to be Abelian  in order for that term to be gauge invariant). 
The ellipses denote other terms involving two or more derivatives (including in principle the Standard Model
fields), 
and $T_b$, $\mathcal{A}_b$, $\mathcal{B}_b$, $\mathcal{C}_b$ 
and so on could depend on the scalars $\phi^i$.

One can now ask again whether the observed value of the cosmological constant emerges naturally in this 
framework. While no positive answer is found in non-supersymmetric cases, the supersymmetric model 
has been proven to have very interesting features \cite{Aghababaie:2003wz,Burgess:2011va}. 
One can compute the cosmological constant in three steps: first, the brane fields are integrated out at the 
quantum level; second, the classical integration in the bulk is performed; third, the quantum corrections
of the bulk integration are computed. The first and second steps always give a vanishing 
cosmological constant if the dilaton does not couple directly to the branes \cite{Aghababaie:2003wz,Aghababaie:2003ar}.
Ref. 
\cite{Burgess:2011va} has recently shown that  the third step is not dangerous either, at least 
for representations of the supersymmetry algebra which are massless in the 6D sense. The essence of the argument
is that  $e^{2\phi}$ acts as a loop counting quantity (as it can be proved by going to the
frame defined by $\hat{g}_{\ssM \ssN} = e^{\phi} g_{\ssM \ssN}$) and is very small, of order
$1/(r^4 \mbox{TeV}^4)$. The 
last property can be understood by noticing that the bulk Lagrangian enjoys a classical scale invariance
under which $e^{\phi}g_{\ssM \ssN}$ does not change; in the large extra dimension setup the value $e^{\phi} r^2$ is therefore expected 
to be fixed by the field equations to a value of order $1/\mbox{TeV}^2$.  The contents of the following
sections are useful tools to address the same question in the presence of massive 6D supermultiplets
(see also \cite{Burgess:2005cg} for a study of bulk UV sensitivity for
Ricci-flat geometries, including the gravity sector, but without branes.).

Let us conclude this section by discussing the solutions of the models we have considered. The simplest solution 
preserving 4D Poincar\'e invariance and providing a flux stabilization is a $(Minkowski)_4\times S^2$ compactification, where the metric of the two extra dimensions is that of a sphere with radius $r$,
\be
 d s^2 = r^2 ( d \theta^2 + \sin^2 \theta \, d\varphi^2 ) \,.
\ee
($0\leq \theta \leq \pi$ and $0\leq \varphi < 2\pi$). Also, the scalars $\phi^i$  are  constant and the non-vanishing components of the field strength are
$F_{mn}=f \epsilon_{mn}$, where $f$ is a constant. If there are some matter fields having a non-trivial charge\footnote{We denote with $\tilde{g}$ the gauge coupling of the background gauge field, while $q$ is the charge of a given field, which we assume to be integer.},
$q\tilde{g}$, under
this background gauge field, then its field strength $F_{mn}$ has to satisfy the following 
quantization condition \cite{RandjbarDaemi:1982hi}
\be \label{eq:fluxqtzn}
2\pi N = q \int_{S_2} F  = 4 \pi r^2 qf \,,  \quad (\mbox{without brane sources})
\ee
where $N = 0, \pm 1, ...$ is an arbitrary integer. Then $f$ must satisfy
\be \label{E:fquant}
    f = \frac{N}{2  q r^2} \,. \quad (\mbox{without brane sources})
\ee
In the supersymmetric model, the field equations also imply
\be \label{E:phircond}
e^{\phi}  = \frac{\kappa^2}{4 g^2\, r^2}  \quad \mbox{and} \quad
 f = \pm \frac{\tilde{g}}{2 \,g \, r^2} \, .  \quad (\mbox{supersymmetric case})
\ee
Since all the parameters of the bulk theory are expected to be of order TeV (to the appropriate power) in the large extra dimensions setup, we notice that the first of these two conditions confirms our general expectation that
$e^{\phi}r^2$ is of order $1/\mbox{TeV}^2$. The second one, combined with Eq. (\ref{E:fquant}), gives us $N=\pm q \tilde{g}/g$. When the background gauge field is along the $U(1)_R$, and so $\tilde{g}=g$,
$N=\pm1$ and this solution
preserves half of the supersymmetries of the bulk theory \cite{SS}. For other embeddings of the background gauge field supersymmetry is instead completely broken.

A simple way to introduce 3-branes in this case is starting from the sphere metric
and demanding  $\varphi$ to have a modified period, $\varphi \sim \varphi +2\pi \alpha$, 
where $\alpha$ is a positive constant \cite{conical}. This procedure introduces two conical singularities, one 
at the north pole and the other one at the south pole, 
with identical defect angles, $\delta = 2\pi (1-\alpha)$. The resulting internal space is called the  
{\it rugby ball}.

The brane action in (\ref{eq:genbraneaction}) can support these conical singularities and the near-brane
boundary conditions imply \cite{localizedflux} that the defect angle is 
\be
 \delta_b = \kappa^2 L_b \,.\label{deltab-Lb}
\ee
In order for this angle to be positive we will assume that $L_b\geq 0$ at the background solution.
The  brane sources also changes the flux quantization condition. This 
 arises because the branes themselves can support a localized flux:
\be
 \Phi_b = \frac{q \mathcal{A}_b}{2\pi} \,.
\ee
If the  branes are identical  the total flux localized in this way is 
$\Phi := \sum_b \Phi_b$, in terms of which the flux-quantization condition is (when the background gauge field  has some localized flux)
\be \label{eq:BLFquantization}
 2 \pi N = q\left(\sum_b \mathcal{A}_b + \int_{S_2(\alpha)} F \right)
 = 2\pi \, \Phi + 4 \pi \alpha\, r^2 q f \,,
\ee
where $S_2(\alpha)$ is the rugby ball.
The normalization constant $f$ is then given by
\be \label{f-with-branes}
f= \frac{\mathcal{N}}{2 q r^2} \,,\quad (\textrm{with brane sources})
\ee
where $\mathcal{N} := \omega(N-\Phi)$ and $\omega:=1/\alpha$.

In the supersymmetric case, the above-mentioned classical scale invariance can be broken by the boundary localized fluxes, the second term in Eq. (\ref{eq:genbraneaction}), which gives mass to the 
corresponding flat direction  \cite{localizedflux}; another reason why these fluxes are useful is that they allow us to recover supersymmetry, when the background gauge
field is along the $U(1)_R$, in a continuous limit \cite{localizedflux,Burgess:2011va}. To understand this let us  take the rugby ball solution, $\tilde{g}=g$ and $\Phi_b=0$, then $\mathcal{N} = \omega N$ and the second condition in 
Eq. (\ref{E:phircond}) together with (\ref{f-with-branes}) implies $\omega N=\pm q$,  which does not allow you to approach the supersymmetric value $\omega=1$ continuously. In the absence of $\Phi_b$  instead we do not find this obstruction: using the second condition of 
 (\ref{E:phircond}) in  (\ref{f-with-branes}) this time we obtain 
\be  \frac{\Phi}{q}=\frac{N}{q}\mp \alpha \frac{\tilde{g}}{g}\, ,  \quad (\mbox{supersymmetric case})\ee
 which, in the case $\tilde{g}=g$ and $N/q=\pm 1$, gives 
 \be   \frac{\Phi}{q}= \pm (1-\alpha)\, . \quad  (\mbox{supersymmetric case})\ee 
In the presence of boundary localized fluxes supersymmetry can be broken by an arbitrarily small amount. For this reason in the supersymmetric model we will always take the boundary localized flux in the $U(1)_R$ direction,
$g_b=g$.

Let us conclude this section by mentioning that, for branes carrying both tension and flux, there is, remarkably, a case in which one half of the 6D supersymmetries is preserved; this  occurs when the background gauge
field is along the $U(1)_R$ generator, for which there is exactly the same amount of localized flux on the two branes  \cite{articleB} (i.e. $\Phi_b=\Phi/2$ both at the north and south branes). Such property has important implications   regarding the smallness of the cosmological constant.

\section{Spectrum in codimension-two brane worlds }\label{spectra}

We now move to the analysis of the linear perturbations around the 3-brane solutions we have considered. Since
these configurations preserve 4D Poncar\'e symmetry, such analysis is equivalent to computing
the 4D particle spectrum (defined in the usual sense). This will provide us with additional physical information; 
it is an important computation  to study the stability of the background solutions \cite{ Parameswaran:2007cb,Parameswaran:2009bt} and, of special relevance
for this article, it is an intermediate step to determine the quantum corrections to the 4D vacuum energy. 
Indeed  in dimensional regularization the one-loop contribution to the quantum potential (the Casimir energy)  due to a generic  field with 
bulk mass $m$ is
\be\label{effective potential}
V=\frac{1}{2}\mu^{4-d}(-1)^{F}\sum_{\underline{n}}  
\int\frac{d^dp}{(2\pi)^d}\ln\left(\frac{p^2 + m^2_{\underline{n}} +m^2}{\mu^2}\right), \label{VfromKK}
\ee
where the collective index $\underline{n}$ includes all  KK numbers and $m_{\underline{n}}$ represents the full set of
KK  masses. Also $(-1)^{F} $ is $1$ for bosons
and $-1$ for fermions. The 4D particle spectrum is therefore an important ingredient to compute $V$ and we will
use Eq. (\ref{VfromKK}) to compute the renormalization of bulk and brane couplings in sections \ref{Casimir-renormalization}
and \ref{Casimir-renormalization-RB}.

In the presence of two extra dimensions we have two KK numbers, $\underline{n}=j,n$; in the rest of this section
we give the form of $m^2_{jn}:=\lambda_{jn}/r^2$ for (minimally coupled) scalars, fermions and gauge fields on top of the rugby ball geometry\footnote{Another analysis of the spectrum in codimension-two brane worlds,  but for other background solutions can be found for example in \cite{Giovannini:2002sb} (see also \cite{Giovannini:2001fh} for preliminary technical aspects).} 
    sourced by branes
with both tension $T_b$ and flux $\Phi_b$ (the leading terms in a derivative expansion). In the absence of localized fluxes, these $m^2_{jn}$ were computed\footnote{Those
works considered a more general compactification \cite{Gibbons:2003di} than the rugby ball one, which involves generically
non-factorizable geometries.
Here, however, we focus on the rugby ball case for the sake of simplicity (see \cite{Parameswaran:2010pd}
for the explicit unwarped limit).}
in Refs. \cite{Parameswaran:2006db,Salvio:2007mb,Parameswaran:2007cb}. Subsequently, their form in the presence of $\Phi_b$ was derived in \cite{articleA}.

Let us consider first the simple case of a minimally coupled real scalar, satisfying the equation  $(D_\ssM D^\ssM +m^2)\phi = 0$, 
that is coupled to the background gauge field through a monopole number $N$ and boundary localized fluxes 
$\Phi_b$. In this case the scalar spectrum (in the north patch of the gauge potential) is 
\be \label{eq:simplescalarspec}
 \lambda^{\rm s}_{jn} = \left(j + \frac{\omega}{2} |n - \Phi_+|+\frac{\omega}{2} |n-N+\Phi_-|
 +\frac{1}{2}\right)^2 - \frac{1 +\mathcal{N}^2}4 \,,
\ee
where $$j=0,1,2,... \,,\quad n=0,\pm 1, \pm 2$$ and   $\Phi_+$ ($\Phi_-$) is the flux localized on the north (south) pole of the rugby ball, where
$\cos\theta=+1$ ($-1$).

Moving to fermions, the KK spectrum for a field, satisfying $(\Dsl+m)\psi =0$, that is charged under the $U(1)$
is (using again the north patch of the gauge potential) 
\be
 \lambda^{\rm f\sigma}_{jn} = \left( j + \frac{\omega}{2}\left|n_{1/2} -\Phi_+ -\frac{\sigma}{2\omega}\right| 
 + \frac{\omega}{2} \left|n_{1/2}-N+\Phi_- +\frac{\sigma}{2\omega} \right| +\frac{1}{2}\right)^2 - \frac{\mathcal{N}^2}{4}\,, \label{eq:simplefermionspec}
\ee
where $n_{1/2} = n-\sigma/2$ and $\sigma \in \{\pm 1\}$ corresponds to the 4D helicity of the spinor, 
of which there are 2 (4)  each for a  6D Weyl (Dirac) spinor. 

Finally let us give the KK spectrum arising from a gauge field, which assume {\it not} to be in the 
Lie algebra direction where the background gauge field lies\footnote{This assumption simplifies our calculation
because the gauge field perturbations along the same Lie algebra direction as the background gauge field mix with 
the metric perturbations. See, however, Ref. \cite{Parameswaran:2009bt}, for a discussion of this more complicated
sector, but in the absence of boundary localized fluxes.}. 
There a two cases: the gauge field
can either be massless or massive. We begin with the massless case, when the field satisfy the equation
$g^{\ssM\ssN}D_\ssM F_{\ssN\ssP} =0$. In an appropriate gauge (e.g. light-cone gauge 
\cite{Randjbar-Daemi:1984ap,Randjbar-Daemi:1984fs,RandjbarDaemi:2002pq,Parameswaran:2007cb,Parameswaran:2009bt})
the 6D gauge field can be decomposed into four components, each with a spectrum 
(once again in the north patch of the gauge potential) given by
\be \label{4D-gaugescalar-spectrum}
\lambda_{jn}^{\rm gf\xi} = \left( j + \frac{\omega}{2} \left| n  - \Phi_+ + \frac{\xi}{\omega} \right| 
+ \frac{\omega}{2} \left| n - N +\Phi_- - \frac{\xi}{\omega} \right| +\frac{1}{2}\right)^2 - \frac{(1+\mathcal{N}^2)}4\,,
\ee
where $\xi \in \{0,0,+ 1, -1\}$ for each of the four components and we assume $N=0,\pm 1$ to ensure stability 
\cite{Randjbar-Daemi:1983bw,Parameswaran:2007cb}. We observe that
two modes have exactly the same spectrum as scalars (i.e.~those with $\xi=0$), 
while the other two (with $\xi = \pm 1$) have almost the same spectrum. 
For massive gauge fields we have a more complicated situation because we need a scalar field $\Phi$ that is charged
under the gauge field we are studying in order to give mass through the Higgs mechanism. 
In order to interpret $\langle \Phi\rangle \neq  0$ as a 6D spontaneous symmetry breaking, we require $\langle \Phi\rangle$
to
be constant and to be  at the minimum of $U$. Then, in order 
to solve the background scalar equation, $D_\ssM D^\ssM \Phi=0$,
we also demand that $\langle \Phi\rangle \neq  0$ does not break the $U(1)$ where the background gauge field lies:
otherwise it would not be  possible to have $\langle \Phi\rangle$  constant, 
at least in the sphere compactification of interest
in this paper \cite{RandjbarDaemi:2006gf,Salvio:2006mh,Salvio:2007mb}.
If we choose again the light-cone gauge, it is possible to show that a massive gauge field leads
to the 4D spectrum of a massless gauge field, Eq. (\ref{4D-gaugescalar-spectrum}), 
plus that of a scalar, Eq. (\ref{eq:simplescalarspec}).

\section{Casimir energy  and bulk renormalization} \label{Casimir-renormalization}

The purpose of this section is to compute the Casimir energy due to bulk loops and  to show how, consequently, the bulk and brane couplings renormalize. 
This is an important step to address any problem of technical naturalness, such as the gauge hierarchy and the cosmological constant
problem. We will consider in some detail the case in which the branes are of codimension-two, but some of our results will be valid
in more general brane worlds. The renormalized couplings will depend as usual on a renormalization energy $\mu$ and we will compute explicitly their RGEs.

\subsection{A general technique to compute the Casimir energy}\label{General-technique}

The starting point of this calculation is  the formula for the one-loop contribution of a single field to the quantum potential in Eq. (\ref{effective potential}). Notice that, modulo terms which are independent of $X$, we have  $\ln X = - \int_0^\infty (d s/s) \exp(-s X)$ and therefore
\be
V=-\,\frac{1}{2} \, \mu^{4-d}(-1)^F \sum_{\underline{n}} \int \frac{d^d p}{(2\pi)^d} \, \int_0^\infty \frac{d s}{s} \, \exp \left[ - s \left(p^2 +m^2_{\underline{n}} +m^2  \right) \right]\, .
\ee 
Performing the integral in $d^dp$ and rescaling the variable $s$ we obtain
\be V= -\frac{\mu^4}{2(2\pi)^d}(-1)^F \sum_{\underline{n}} \int_0^{\infty}\frac{dt}{t^{1+d/2}} e^{-\pi t \left(m^2_{\underline{n}} +m^2\right)/\mu^2} \label{VDR}. \ee 
Consider first the case Re$(d)<0$ and  rewrite the integral in 
Eq. (\ref{VDR}) as 
\be \int_0^{\infty}\frac{dt}{t^{1+d/2}} e^{-\pi t\, (m^2_{\underline{n}} +m^2)/\mu^2} = \left(\frac{\pi m^2_{\underline{n}} +\pi m^2}{\mu^2 }\right)^{d/2} 
\Gamma\left(-d/2\right)\, . \ee
We now extend this integral function by analytic continuation to all complex $d$ except the non-positive integers 
(where the function has simple poles). By using 
\be \Gamma\left(-d/2\right)= -\frac{1}{d-4} +\frac{3}{4} -\frac{\gamma}{2} + \mathcal{O}\left(d-4\right)\,, \ee 
where $\gamma$ is Euler's constant, one then obtains
\be V= -\frac{1}{2 \left(4\pi\right)^2r^4}(-1)^F\sum_{\underline{n}} \,( r^2m^2_{\underline{n}} +r^2 m^2)^{d/2} \left( -\frac{1}{d-4} + \ln(r \mu) + \, .... \right)\, ,
\label{Vd}
\ee
where the dots represent finite $r$-independent terms. 

One possible renormalization scheme (which we will adopt from now on) is to subtract the divergent part in the brackets of Eq. (\ref{Vd}). The renormalized potential $V_r$ (the Casimir energy) can then be written as 
\be 
V_r= \frac{C}{(4 \pi )^2r^4}\log\left( \frac{r}{r_0}\right)\, ,\label{1loopV}
\ee
where $r_0^{-1}$ has to be identified as a ultraviolet (UV) scale which can be computed once the UV completion is known and 
\be C:= -\frac{1}{2 }(-1)^F\sum_{\underline{n}} \,( r^2m^2_{\underline{n}} +r^2 m^2)^{d/2}\, .\label{C}\ee
It is important to notice that all we need in order to compute this coefficient is the divergent part of $V$, as it is clear from Eq. (\ref{Vd}).

The quantity on the right hand side of Eq. (\ref{C}) is divergent and has to be regularized. Notice that the exponent $d/2$ can effectively act as a regulator for the sum over $\underline{n}$ (zeta function regularization \cite{Hawking:1976ja}): 
for $d= 4$ the sum is divergent, but one can (and we will) compute (\ref{C}) for those $d$ such that the sum is convergent and then consider the analytic continuation of the final result at $d=4$. 

\subsection{The Casimir energy for rugby ball compactifications} \label{zeta-function-DR}

A generic form of the spectrum which covers all the cases encountered here, Eqs. (\ref{eq:simplescalarspec}),
(\ref{eq:simplefermionspec}) and 
(\ref{4D-gaugescalar-spectrum}), is
\be m^2_{jn}:=\frac{1}{r^2}\lambda_{jn}\, , \qquad \mbox{where} \quad  \lambda_{jn}= \left(j + \frac{\omega}{2} \left| n + b_+ \right| + \frac{\omega}{2} \left| n - b_- \right| + a \right)^2 - \tau \,, \label{generic-spectrum}\ee
and $b_{\pm}$, $a$ and $\tau$ are real parameters; the only assumption we make is that $b_{\pm}$ and $a$ are independent of $j$ and $\tau$ is independent of both $j$ and $n$.

Then the contribution to the $C$ parameter from a single field is 
\be C= -\frac{(-1)^F}{2} \sum_{n, j}  \left\{\left(j + \frac{\omega}{2} \left| n + b_+ \right| + \frac{\omega}{2} \left| n - b_- \right| + a \right)^2 - \tau +(mr)^2\right\}^{d/2}.\ee
where $(-1)^F=1$ for a boson and $(-1)^F=-1$ for a fermion. We can now expand $\left\{...\right\}^{d/2}$ by using 
the binomial series to obtain
\bea  C= -\frac{(-1)^F}{2}  \sum_{n, j} \sum_{k=0}^{\infty}\frac{ \Gamma(k-d/2)}{k! \Gamma(-d/2)}\left[j + \frac{\omega}{2} \left| n + b_+ \right| + \frac{\omega}{2} \left| n - b_- \right| + a\right]^{d-2k} \left[ \tau -(mr)^2\right]^k. 
\label{binomial}\eea 
The sum over $j$ can be performed by means of   the following representation of the Hurwitz zeta function $\zeta (s,c)$  (valid for Re$(s)<0$ and Re$(c)>0$) 
 \be \zeta (-s,c) := \sum_{j=0}^{\infty} (j+c)^s =\frac{1}{\Gamma[-s]}\int_0^{\infty}dy \,\,\frac{y^{-1-s}\, 
 e^{-c y}}{1-e^{-y}}. \label{zetasa}
\ee 
Indeed setting $s=d-2k$ and $c=c_n=  ( \left| n + b_+ \right| +\left| n - b_- \right|) \omega /2+a$ in (\ref{zetasa}) we have
 \be  C= -\frac{(-1)^F}{2} \sum_{k=0}^{\infty} \frac{ \Gamma(k-d/2) \left[\tau -(mr)^2\right]^k}{k! \Gamma(-d/2)\Gamma(2k-d)} \int_0^{\infty}dy \,\,\frac{y^{-1+2k-d}\, 
}{1-e^{-y}} \sum_n e^{-c_n y}.  \label{passaggio}\ee
Observing that  integrals of the form $ \int_0^{\infty} dy \frac{ y^h e^{-y}}{1-e^{-y} }$ can be computed explicitly
for Re$(h) > 0$,
\be  \int_0^{\infty} dy \frac{ y^h e^{-y}}{1-e^{-y} }=\Gamma[1 + h]\mbox{Li}_ {1+h}(1), \qquad \mbox{Li}_ {1+h}(1):=  \sum_{k=1}^{\infty}\frac{1}{k^{1+h}},\label{Integral}\ee
we Taylor-expand  $ye^{y}\sum_n e^{-c_n y}$ around $y=0$,
\be ye^{y}\sum_n e^{-c_n y} = \sum_{k'=0}^{\infty} \frac{ \epsilon_{k'}(\omega) }{k'!}y^{k'}, \label{expansion1}\ee
and obtain 
\be C= -\frac{(-1)^F}{2}\sum_{k=0}^{\infty} \frac{ \Gamma(k-d/2) \left[\tau -(mr)^2\right]^k}{k! \Gamma(-d/2)\Gamma(2k-d)} \sum_{k'=0}^{\infty} \frac{ \epsilon_{k'}(\omega) }{k'!}\Gamma(-1-d+2k+k')\mbox{Li}_ {-1-d+2k+k'}(1). \label{C1gauge}\ee
We now need to take the limit $d\rightarrow 4$. The previous expression turns out to be well defined in this  limit. Also, because of the $\Gamma(-d/2)$ in the denominator, only a finite number of $k$ and $k'$ contributes  to the sum: $k=0,1,2,3$ and 
$k'=0,1,2,3,4,5,6$. This implies two things: $(i)$ $C$ is computable once we know  $\sum_n e^{-c_n y}$, $(ii)$ $C$ has the generic structure
\be 
C = \frac{s_{-1}}{6} (m r)^6 - \frac{s_0}{2} (m r)^4 + s_1 (m r)^2 - s_2 \,. \label{Cm}
\ee

Since the coefficient $C$ can be computed from the mere knowledge of the UV divergent part of the Casimir energy, we should expect that this formula can be applied to codimension-two compactifications which 
are more general than the rugby ball one. This is because UV divergences are related to the local structure of 
the space-time and are therefore insensitive to global properties such as its topology.

\subsection{Bulk and brane counterterms and renormalized couplings}

The models we have considered are non-renormalizable and so not all divergences can be reabsorbed in counterterms of the same form as the terms in the classical action. However, one can show in very general terms that the number of  counterterms needed is always finite at a given order in perturbation theory. In this subsection we review \cite{articleA} the  renormalization due to bulk loops in codimension-two braneworlds (and in particular for rugby ball compactifications) at the one-loop level. At the end we will therefore obtain a finite number of counterterms and renormalized couplings.
To perform this calculation we will use effective field theory methods (see for example \cite{Burgess:2003jk}).

Although only bulk loops\footnote{Loops of brane localized fields can be calculated and the corresponding divergences can be reabsorbed with the usual 4D  field theory methods.} are computed here, both bulk ad brane counterterms are needed.  There is an important difference between them. 
The bulk counterterms, unlike the brane ones, do not depend on the brane properties and so they can be computed in
the sphere limit, $\alpha \rightarrow 1$.

\subsubsection{Renormalization of the bulk interactions.} To capture all the terms needed to reabsorb the 
UV divergences we write down the most general local Lagrangian with the chosen field content and set of symmetries, which we organize in a derivative expansion $\mathcal{L}_\ssB^{\rm ct}=\mathcal{L}^{\rm ct}_{\ssB 0}+\mathcal{L}^{\rm ct}_{\ssB 2}+\mathcal{L}^{\rm ct}_{\ssB 4}$. After renormalization this generates a corresponding series of renormalized interactions $\mathcal{L}_\ssB^{r}=\mathcal{L}^r_{\ssB 0}+\mathcal{L}^r_{\ssB 2}+\mathcal{L}^r_{\ssB 4}$. Focusing on the fields that are non-zero in the background we have 
\bea
\mathcal{L}^{r}_{\ssB 0} &=&  - \sqrt{-g} \;  \lambda \, , \nonumber\\
 \mathcal{L}^{r}_{\ssB 2} &=&  - \sqrt{-g} \left[ \frac{\zeta_\ssR }{2 \kappa^2} \, R + \frac{\zeta_\ssA}{4\tilde{g}^2} \, F_{\ssM \ssN} F^{\ssM \ssN} \right] \, ,\\
 \mathcal{L}^{r}_{\ssB 4} &=& - \sqrt{-g} \left[ \frac{ \kappa \zeta_{\ssA\ssR}}{8 \tilde{g}^2} \,  R \, F_{\ssM \ssN} F^{\ssM \ssN} + \frac{\zeta_{\ssR^2}}{\kappa} \, \bar{R}^2 \right] \, , \\
\mathcal{L}^{r}_{\ssB 6} &=& - \sqrt{-g} \left[\zeta_{\ssR^3}\, \bar{R}^3 + ...\right]\, , \\ 
&...& \, ,
\eea
where $\bar{R}^2$ ($\bar{R}^3$) is a generic linear combination of terms which are quadratic (cubic) in the curvature, that is 
\be \label{Rsqol}
\bar{R}^2 = a_\ssR\, R^2 + 2 b_\ssR\, R_{\ssM \ssN} R^{\ssM \ssN} + c_\ssR\, R_{\ssM \ssN \ssP \ssQ} R^{\ssM \ssN \ssP \ssQ} 
\ee
where $a_\ssR + b_\ssR + c_\ssR = 1$ so that $\bar{R}^2 = R^2$ when specialized to the sphere geometry (for which $R_{mnpq} R^{mnpq} = 2 R_{mn} R^{mn} = R^2 = 4/r^4$). A similar expression is used for $\bar{R}^3$. Calculations on a sphere can only provide the overall couplings $\zeta_{\ssR^2}$, $\zeta_{\ssR^3}$ and not the separate parameters such as $a_\ssR$, $b_\ssR$ and $c_\ssR$ (see however \cite{articleA} and references 
therein to know the latter quantities).

Evaluating the renormalized action at the background sphere solution and integrating over the extra dimensions gives
\bea \label{VbulkAtEom}
 V_\ssB^{r} = - \int d^2 x \, \mathcal{L}_\ssB^{r}&=& \left( 4 \pi  r^2 \right) \left\{ \lambda - \frac{\zeta_\ssR}{\kappa^2 r^2} + \frac{f^2}{2 \tilde{g}^2} \left[ \zeta_\ssA - \frac{\kappa \zeta_{\ssA\ssR} }{r^2} \right]  + \frac{4 \zeta_{\ssR^2}}{\kappa \,r^4}  - \frac{8\zeta_{\ssR^3}}{r^6} +... \right\} \nonumber \\
 &=& \left( 4 \pi  r^2 \right) \left\{ \lambda - \frac{\zeta_\ssR}{\kappa^2 r^2} + \frac{\mathcal{N}^2}{8 \,q^2 \tilde{g}^2 r^4} \left[ \zeta_\ssA - \frac{\kappa \zeta_{\ssA\ssR} }{r^2} \right]  + \frac{4 \zeta_{\ssR^2}}{\kappa \,r^4} - \frac{8\zeta_{\ssR^3}}{r^6} +... \right\}\!\!.
\eea
Therefore $\lambda$, $\zeta_\ssR$, $\zeta_{\ssR^2}$, and $\zeta_{\ssR^3}$   can be read off respectively from the $r^2$, $r^0$, $r^{-2}$, and $r^{-4}$ terms in $V_r$ (see Eqs. (\ref{1loopV}) and (\ref{Cm})), while the $\zeta_\ssA$ and $\zeta_{\ssA\ssR}$ coefficients are identified as the $\mathcal{N}^2/r^2$ and $\mathcal{N}^2/r^4$ terms respectively. 
This implies that integrating a bulk field with mass $m$ gives the following contribution to the RGEs
\bea
 \mu \, \frac{\partial \lambda}{\partial \mu} = \frac{m^6}{6 (4\pi)^3} \; s_{-1}^{\rm sph,\, 0} \,,&&\quad
 \mu \, \frac{\partial}{\partial \mu} \left( \frac{\zeta_\ssR}{\kappa^2} \right) = \frac{ m^4}{2 (4\pi)^3} \; s_{0}^{\rm sph,\,0} \,, \label{RGEs-bulk}\\
 \mu\,\frac{\partial }{\partial \mu}\left(\frac{\zeta_{\ssR^2}}\kappa\right) = \frac{m^2}{4(4\pi)^3} \; s_1^{\rm sph,\,0} \,,&&\quad
\quad  \mu\,\frac{\partial \zeta_{\ssR^3}}{\partial\mu} = \frac{1}{8(4\pi)^3} \;s_2^{\rm sph,\,0} \,.
\eea
\be
 \mu \, \frac{\partial}{\partial \mu} \left(\frac{\zeta_\ssA}{\tilde{g}^2}\right) = \frac{2\, m^2}{(4\pi)^3 r^4 f^2} \; s_{1}^{\rm sph,\,2} = \frac{8\, q^2 m^2}{(4\pi)^3 \mathcal{N}^2} \; s_{1}^{\rm sph,\,2} 
\ee
and
\be
 \quad \mu \, \frac{\partial}{\partial \mu} \left(\frac{\kappa \zeta_{\ssA\ssR}}{\tilde{g}^2}\right) = \frac{2}{(4\pi)^3 r^4 f^2} \; s_{2}^{\rm sph,\,2} = \frac{8 \, q^2}{(4\pi)^3 \mathcal{N}^2} \; s_{2}^{\rm sph,\,2}  \label{RGEs-bulk2}\
\ee
and so on. The ``sph'' in  $s_i^{{\rm sph},\,k}$  emphasizes that these quantities are evaluated on the sphere, while the superscript `$k$' denotes terms involving $k$ powers of $\mathcal{N}$.  
The renormalization of the gauge-field terms, $\zeta_\ssA$ and $\zeta_{\ssA\ssR}$, has been done by looking at the $\mathcal{N}$-dependent divergences produced when a particle with charge $q \tilde{g}$ runs in the loop.

\subsubsection{Renormalization of the brane interactions.} In this case we have a dependence on the boundary conditions used near the brane but the result should be independent of the boundary conditions on distant branes. 

To compute the brane contributions we first subtract the (boundary condition independent) bulk contributions found above. Noticing that the bulk counterterms should be integrated over the volume of the rugby ball, which is $4\pi \alpha\, r^2$, we  define
\be \label{eq:deltasidef}
 \delta s^{\rm tot}_i = s_i - \alpha \, s_i^{\rm sph} \,,
\ee
and use $\delta s^{\rm tot}_i = \sum_b \delta s_{i(b)}$ to extract how the interactions on each individual brane renormalize. This can be done as before, by distinguishing the interactions that depend on the gauge field which is non-zero on the background from those that do not. In the following we understand the label $(b)$ in $ \delta s_{i(b)}$ to have a simpler notation.

Writing the most general local brane Lagrangian organized  in a derivative expansion,  $\mathcal{L}^{r}_b = \mathcal{L}^{r}_{b0} +\mathcal{L}^{r}_{b1} +\mathcal{L}^{r}_{b2} +\mathcal{L}^{\rm ct}_{b3} + ...$, and dropping terms that vanish at  the background, we have
\be
 \mathcal{L}^{r}_{b 0} =  - \sqrt{-\gamma} \;  T_b \, ,
\ee
\be
\mathcal{L}^{r}_{b 1} =  \sqrt{-\gamma} \left[ \frac{ \zeta_{\tilde{\ssA} b}}{2\tilde{g}^2} \, \epsilon^{mn} F_{mn} \right]\, ,
\ee
\be
\mathcal{L}^{r}_{b 2} =  - \sqrt{-\gamma} \left[ \frac{
 \zeta_{\ssR \,b} }{\kappa} \, R + \frac{\kappa \zeta_{\ssA b}}{4 \tilde{g}^2} \, F_{\ssM \ssN} F^{\ssM \ssN} \right]\, ,
\ee
\be
\mathcal{L}^{r}_{b 3} = \sqrt{- \gamma} \left[ \frac{\kappa \zeta_{\tilde{\ssA} \ssR \, b}}{2 \tilde{g}^2} \,  R \, \epsilon^{mn} F_{mn} \right] \,,
\ee
\be
\mathcal{L}^{r}_{b 4} = - \sqrt{-\gamma} \left[ \zeta_{\ssR^2 b} \, \bar{R}^2 + \frac{\kappa^2 \, \zeta_{\ssA \ssR \,b}}{8 \tilde{g}^2} \, R \, F_{\ssM \ssN} F^{\ssM \ssN} \right]  \,,
\ee
and so on, where $\gamma_{\mu\nu} := g_{\ssM \ssN} \partial_\mu x^\ssM \partial_\nu x^\ssN$ (with 
the right-hand side computed at the brane position) is the induced metric on the brane. 

Evaluating these at the background solution gives the following contribution to the Casimir energy
\bea
 V_b^{r} &=&  T_b - \frac{\zeta_{\tilde{\ssA}b} f}{\tilde{g}^2} - \frac{2 \zeta_{\ssR \, b}}{\kappa \,r^2} + \frac{\kappa \zeta_{\ssA b} f^2}{2 \tilde{g}^2} + \frac{2 \, \kappa \zeta_{\tilde{\ssA}\ssR \, b} f}{\tilde{g}^2 r^2} + \frac{4 \zeta_{\ssR^2 b}}{r^4} - \frac{\kappa^2\zeta_{\ssA \ssR \, b} f^2}{2 \, \tilde{g}^2 r^2} +... \nonumber\\
 &=&  T_b - \frac{\zeta_{\tilde{\ssA}b} \mathcal{N}}{2 \, q \tilde{g}^2 \,r^2} - \frac{2 \zeta_{\ssR \, b}}{\kappa \,r^2} + \frac{\kappa \zeta_{\ssA b} \mathcal{N}^2}{8 \, q^2 \tilde{g}^2 r^4} + \frac{ \kappa \zeta_{\tilde{\ssA}\ssR \, b} \mathcal{N}}{q \tilde{g}^2 r^4} + \frac{4 \zeta_{\ssR^2 b}}{r^4} - \frac{\kappa^2 \zeta_{\ssA \ssR \, b} \mathcal{N}^2}{8 \,  q^2\tilde g^2 r^6} +... \,.
\eea
Using Eqs. (\ref{1loopV}) and (\ref{Cm}) then gives the following RGEs
\bea
 \mu \, \frac{\partial \, T_b}{\partial \mu} = -\frac{m^4}{2(4\pi)^2} \; \delta s^0_{0}\,,\quad&& \mu\,\frac{\partial}{\partial \mu} \left( \frac{\zeta_{\tilde{\ssA} b}}{\tilde{g}^2}\right) = - \frac{2\,q m^2}{(4\pi)^2 \mathcal{N}} \; \delta s^{1}_1 \,, \nonumber\\
 \mu\,\frac{\partial}{\partial \mu} \left( \frac{\zeta_{\ssR b}}{\kappa}\right) = - \frac{m^2}{2(4\pi)^2} \; \delta s^0_1 \,,\quad&& \mu\,\frac{\partial}{\partial \mu} \left( \frac{\kappa \zeta_{\tilde{\ssA} \ssR \, b}}{\tilde{g}^2}\right) = -\frac{q}{(4\pi)^2 \mathcal{N}} \; \delta s^{1}_{2} \,, \label{RGEs-brane}\\
 \mu\,\frac{\partial \zeta_{\ssR^2 b}}{\partial \mu} = -\frac1{4(4\pi)^2}\; \delta s_2^0 \,,\quad&& \mu\,\frac{\partial}{\partial \mu} \left( \frac{\kappa \zeta_{\ssA b}}{\tilde{g}^2}\right) = - \frac{8\, q^2}{(4\pi)^2 \mathcal{N}^2} \; \delta s^{2}_{2} \,,\nonumber
\eea
where $\delta s_2^k$ are terms with $k$ powers of $\mathcal{N}$.

\section{ Casimir energy  for rugby balls in explicit cases} \label{Casimir-renormalization-RB}

In this section we apply the method of sections \ref{General-technique} and \ref{zeta-function-DR}  
to compute the Casimir energy, i.e. the coefficients $s_i$, produced by specific matter fields, for rugby ball compactifications 
sourced by branes with tension and flux. Indeed the $s_i$ is all we need to obtain the renormalization of the bulk ad brane couplings, as it is clear from Eqs. (\ref{RGEs-bulk})-(\ref{RGEs-bulk2}) and (\ref{RGEs-brane}). Moreover, as we shall comment later on, the $s_i$ can  be used to extract the 4D cosmological constant \cite{articleA,articleB}.

\subsection{A single real  scalar.} We observe that the method of section \ref{zeta-function-DR} can be applied in the case of real scalars because the spectrum in (\ref{eq:simplescalarspec}) has the form (\ref{generic-spectrum}) with 
\be b_+= -\Phi_+\,,\quad b_-= N-\Phi_- \,, \quad a=\frac12 \,,\quad \tau =\frac{1+\mathcal{N}^2}{4}. \ee
The (renormalized) Casimir energy produced by a real scalar is given by Eq. (\ref{1loopV}) with C given in Eq. (\ref{Cm}) and one obtains the following 
$s_i$ coefficients:
\bea \label{eq:simplescalars1}
s^{\rm s}_{-1} &=& \frac{1}{\omega} \,, \label{s-1scalar}\\
s^{\rm s}_0(\omega,N,\Phi_b) &=& \frac{1}{\omega} \left[ \frac16 + \frac{\omega^2}{6}(1-3F) \right] \,, \\
s^{\rm s}_1(\omega,N,\Phi_b) &=& \frac{1}{\omega}\left[ \frac{1}{180} - \frac{\mathcal{N}^2}{24} + \frac{\omega^2}{18}(1 - 3 F) -\frac{\omega^3\mathcal{N}}{12} \sum_b \Phi_b \, G_b + \frac{\omega^4}{180} (1 -15F^{(2)}) \right] ,\qquad \\
s^{\rm s}_2(\omega,N,\Phi_b) &=& \frac{1}{\omega} \left[ -\frac1{504} - \frac{11\,\mathcal{N}^2}{720} + \left(\frac{1}{90} -\frac{\mathcal{N}^2}{144} \right) (1-3F) \omega^2 -\frac{\omega^3\mathcal{N}}{24} \sum_b \Phi_b \, G_b \right.  \nonumber\\
&& \qquad  + \frac{\omega^4(1-\mathcal{N}^2)}{360}(1 - 15F^{(2)}) - \frac{\omega^5\mathcal{N}}{120}\sum_b \Phi_b \,G_b(1+3F_b) \nonumber   \\
&&\qquad   + \left(\frac1{1260} -\frac{F^{(2)}}{120} - \frac{F^{(3)}}{60} \Bigg)\omega^6 \right] \label{eq:simplescalars2} \,, \label{s2scalar}
\eea
where we introduced the notation
\be 
F_b:=|\Phi_b|\left(1-|\Phi_b|\right)\,,  \quad G_b:=\left(1-|\Phi_b|\right)\left(1-2|\Phi_b|\right) \,, \quad F^{(n)}:= \sum_b F_b^n \,.\quad F^{(1)}:=F \,.
\ee

The method of sections \ref{General-technique} and  \ref{zeta-function-DR} can also be applied to fermions and  gauge fields as their spectra, Eqs. (\ref{eq:simplefermionspec}) and (\ref{4D-gaugescalar-spectrum}), are both of the form given in (\ref{generic-spectrum}). The explicit expressions for the $s_i$ coefficients for fermions and 
(massive) gauge fields can be found in \cite{articleA}.

\subsection{Supermultiplets}

Let us now consider supermultiplets of 6D gauged chiral supergravity focusing on the hypermultiplets and the gauge multiplets (for which we further restrict to the case in which the gauge field is zero in the background). The main reason is that the cancellation of gauge and gravitational anomalies typically require hundreds of such supermultiplets \cite{RandjbarDaemi:1985wc,other-anomaly-free}
and so their contribution is expected to dominate the Casimir energy. 
For example, the first anomaly free theory of this sort that has been
 found has a large ($E_6\times E_7 \times U(1)_R$) gauge symmetry
with many (456) hypermultiplets \cite{RandjbarDaemi:1985wc,RandjbarDaemi:2004qr}.

A massless hypermultiplet consists of four massless scalars (called hyperscalars) and one 6D Weyl fermion, the hyperino. 
A massless gauge multiplets is made of one gauge field and a 6D Weyl fermion, the gaugino.  By contrast, a massive
6D matter multiplet consists of a massive gauge field, a massive Dirac fermion and three scalars, a total of eight bosonic and eight fermionic states.
Since this is also the number of degrees of freedom of a gauge plus a hypermultiplet, one expects to form a massive supermultiplet by having the gauge boson from a gauge multiplet `eat' one of the hyperscalars through the Higgs mechanism. 

\subsubsection{Non-supersymmetric embedding of the background gauge field.}
We first consider the case in which the background gauge field is {\it not} embedded in the $U(1)_R$. In this case supersymmetry 
is broken both by the branes and the bulk solution. Since the gauge field whose flux is localized on the branes is 
the $U(1)_R$ gauge field, the spectrum and the Casimir energy as well as the renormalization will not depend on $\Phi_b$ 
for this choice of the gauge field embedding.

The contribution of a hypermultiplet to the $s_i$ coefficients is obtained  by summing the result for a 6D Weyl fermion
to that produced by four hyperscalars\footnote{The hyerscalar contribution is given by the scalar result after
substituting $m^2\rightarrow m^2+1/(2r^2)$ \cite{articleB}.}. We obtain
\bea 
s^{\rm hm}_{-1}(\omega,N) &=& 0\,,\\
s^{\rm hm}_0(\omega,N) &=& \frac{1}{\omega}(-1+\omega^2)\,, \\
s^{\rm hm}_1(\omega,N) &=&  \frac{1}{\omega} \Bigg[ \frac{5}{24}  -\frac{\omega^2}{12}+\frac{\omega^4}{24}- \frac{\omega^2N^2}{2}\Bigg]\,,  \\
s^{\rm hm}_2(\omega,N) &=& \frac{1}{\omega} \left[ -\frac{23}{1440}+\frac{31\, \omega^2}{1440} +\frac{7\, \omega^4}{1440}
+\frac{\omega^6}{160}+\omega^2N^2\left(-\frac{1}{48}-\frac{\omega^2}{24}-\frac{\omega^4}{48}\right) \right] \label{eq:simplescalarshm} \,,\nonumber
\eea
where here $N$ is the common monopole number of the hyperscalars and hyperino.
For a massless ($m=0$) hypermultiplet the only coefficient which matters is $s_2^{\rm hm}$ and the corresponding Casimir energy is 
$-s_2^{\rm hm}\ln(r/r_0)/(4\pi r^2)^2$.

As far as the gauge multiplet is concerned, we should sum the contribution of a gauge field to that of a 6D Weyl fermion,
obtaining for the $s_i$ coefficients 
\bea 
s^{\rm gm}_{-1}(\omega,N) &=& 0\,,\\
s^{\rm gm}_0(\omega,N) &=&  \frac{1}{\omega}\left(1-2\, \omega+\omega^2\right)\,, \\
s^{\rm gm}_1(\omega,N) &=&  \frac{1}{\omega} \left[ \frac{1}{24}  +\frac{\omega^2}{4}+\frac{\omega^4}{24}
+ \frac{\omega^2N^2}{2}\right]\,,  \\
s^{\rm gm}_2(\omega,N) &=& \frac{1}{\omega} \left[ -\frac{7}{1440}+\frac{71\, \omega^2}{1440} +\frac{23\, \omega^4}{1440}
+\frac{\omega^6}{160}+\omega^2N^2\left(-\frac{5}{48}-\frac{\omega}{4}-\frac{\omega^2}{24}-\frac{\omega^4}{48}\right) \right] \label{eq:simplescalarsgm} \,,
\eea
where now $N$ is the common monopole number of the gauge field and gaugino.
For a massless gauge multiplet the only important coefficient  is $s_2^{\rm gm}$ and the corresponding Casimir energy is 
$-s_2^{\rm gm}\ln(r/r_0)/(4\pi r^2)^2$.

Therefore, if one has massless supermultiplets only, the total contribution to the Casimir energy is approximately given
by $-(s_2^{\rm hm}+s_2^{\rm gm})\ln(r/r_0)/(4\pi r^2)^2$. 
The overall sign depends on the particular anomaly free model \cite{RandjbarDaemi:1985wc} that one chooses.

For a massive multiplet made of a hypermultiplet and a gauge multiplet we have that the $s_i$ coefficients are 
$s_i^{\rm mm}= s_i^{\rm hm}+s_i^{\rm gm}$. Therefore, by using the explicit expressions for $s_i^{\rm hm}$
and $s_i^{\rm gm}$  given before, we obtain
\bea 
s^{\rm mm}_{-1}(\omega,N) &=& 0\,,\\
s^{\rm mm}_0(\omega,N) &=&  2(\omega-1)\,, \\
s^{\rm mm}_1(\omega,N) &=&  \frac{1}{\omega} \left( \frac{1}{4}  +\frac{\omega^2}{6}+\frac{\omega^4}{12}
\right)\,,  \\
s^{\rm mm}_2(\omega,N) &=& \frac{1}{\omega} \left[ -\frac{1}{48}+\frac{17\, \omega^2}{240} +\frac{\omega^4}{48}
+\frac{\omega^6}{80}+\omega^2N^2\left(-\frac{1}{8}-\frac{\omega}{4}-\frac{\omega^2}{12}-\frac{\omega^4}{24}\right) \right]  \,.
\eea
From this result, and from Eqs. (\ref{Cm}) and (\ref{1loopV}), we note that for $mr$ at most of order 1
the obtained Casimir energy is $\sim \ln(r/r_0)/(4\pi r^2)^2$. However, for  $mr \gg 1$, integrating out a massive
multiplet gives a dangerously large contribution. We will see that an extra suppression  
can be obtained when the 
gauge background is along the R-symmetry generator.

\subsubsection{Supersymmetric embedding of the background gauge field. } Let us now turn to the case in which the background gauge field is along the $U(1)_R$.  Hyper and gauge multiplet contributions to the Casimir energy have been computed in \cite{articleB} and are quite involved. We therefore refer to this work for explicit expressions associated with  hyper, gauge and massive multiplets. 

One important point we want to emphasize here is that   $s_i^{\rm sph}=0$ for this background gauge field embedding and therefore the renormalization group equation of the bulk Casimir energy vanishes, $\mu\partial_\mu V_\ssB^r=0$. The reason is that, as we mentioned, one half of bulk supersymmetry is not broken in this case.

In the particular case of identical boundary localized fluxes ($\Phi_b=\Phi/2$) also the branes preserve one half of the 6D supersymmetries (see the discussion at the end of section \ref{class}). This implies that for balanced fluxes on the two branes also the brane Casimir energy does not run, $\mu\partial_\mu V_b^r=0.$

When the boundary localized fluxes are unbalanced a non-trivial result  arises. However, by continuity the result should be suppressed by the difference of the two localized fluxes, $\Delta \Phi$. Also, as a remnant of 6D supersymmetry the coefficient $s_{-1}$ for massive supermultiplets vanishes. When 
the bulk mass is such that $mr$ is at most of order one the Casimir energy is $\sim \ln(r/r_0)/(4\pi r^2)^2$. When $mr\gg1$ the contribution of a massive supermultiplet to the Casimir energy is of order $\Delta \Phi  (mr)^4\ln(r/r_0)/(4\pi r^2)^2$ and can be again as small as $\sim \ln(r/r_0)/(4\pi r^2)^2$ for $\Delta \Phi$ appropriately small. 
It is very interesting to notice that the localized flux difference does not receive quantum corrections 
from loops involving brane localized fields only (when they are not charged under the corresponding gauge symmetry)
and therefore taking this difference to be very small 
does not need to have the usual fine tuning.

Let us conclude this section by mentioning that the Casimir energy is not exactly equal to 
 the 4D cosmological constant, but is rather equal to the quantum action evaluated at the classical solution.  
The backreaction of the brane on the bulk is important in this case and leads to a sizeble correction of the solution.
This point is extensively discussed in \cite{articleB}. After taking into account this effect, however,  
the prediction for the (most UV sensitive part of the) 4D cosmological constant can remain as small as $\ln(r/r_0)/(4\pi r^2)^2$.

\section{Conclusions and outlook}

We discussed the Casimir energy as well as the renormalization of bulk and brane  coefficients in the quantum action produced by integrating out (massive) bulk matter in codimension-two brane worlds.  In the  calculation we focused on the 
one-loop approximation and 
explicitly summed over the KK towers. Much of what we presented here  is a review of \cite{Parameswaran:2006db,Parameswaran:2007cb,Parameswaran:2009bt,articleA,articleB}, but we also provided some new results, in particular in the technique to perform the KK sums.

Regarding the motivations,  as discussed extensively in the introduction, codimension-two brane worlds may provide a framework to solve the gauge hierarchy and, when the bulk is supersymmetric, the cosmological constant problem. These are technical naturalness  problems  and as such they require sistematic ways of computing
quantum corrections.

We considered in some detail the rugby ball compactifications in which the size $r$ of the extra dimensions  is stabilized by the flux of a bulk field. A class of model having these configurations as solutions has been presented, including a concrete supersymmetric theory, 6D gauged chiral supergravity. The two 3-branes required to support these solutions can not only carry tension, but also a localized flux of the same gauge field which stabilizes the extra dimensions (tensions and localized fluxes are the leading terms in a derivative expansion of the brane Lagrangians, Eq. (\ref{eq:genbraneaction})).

As discussed in \cite{articleB}, when the localized fluxes on the two 3-branes are identical one half of 
the bulk supersymmetries is unbroken, which implies that the Casimir energy
and, remarkably, the 4D cosmological constant vanish.  

For the rugby ball compactifications the KK spectra of many types of bulk fields are known \cite{Parameswaran:2006db,Parameswaran:2007cb,Parameswaran:2009bt,articleA,articleB} and we reviewed their structure for scalars, fermions and gauge fields (see Eqs. (\ref{eq:simplescalarspec}), (\ref{eq:simplefermionspec}) and (\ref{4D-gaugescalar-spectrum}) rescpectively). 
The explicit form of the spectra allows us to have  explicit expressions for the Casimir energy and, consequently,
for the renormalization group equations, 
which depend on the flux and brane tensions and fluxes. The calculation of the Casimir energy starting from the KK spectrum exploits a novel efficient technique (see sections \ref{General-technique} and \ref{zeta-function-DR}) which allows us to confirm the results of \cite{articleA}.

The explicit form for the Casimir energy we obtain is a polynomial function of $mr$ of degree six, see eqs. (\ref{1loopV}) and (\ref{Cm}). The coefficients $s_i$  of the polynomial are computed explicitly as a function of the brane tensions and bulk and brane fluxes; for example, in the simple case of a bulk scalar they are given in Eqs. (\ref{s-1scalar})-(\ref{s2scalar}).

When the bulk particle that is integrated out is massless, or $m$ is at most of order $1/r$, the 4D cosmological constant has the desired 
order of magnitude regardless of the fact that there is bulk supersymmetry. We show that 
for large $m$ the final result can be appropriately suppressed if we select the supersymmetric model,
the bulk solution preserves
one of the supersymetries and localized fluxes have very similar 
values, such that we are close to a supersymmetric setup.

Let us mention some outlook of the results presented here. 

A possible extension of our work is the inclusion of warped geometries, which represent the most general solutions 
with 4D maximal symmetry. A first step towards this goal is the codimension-one case, which would be interesting
by itself as the Randall-Sundrum model can address the gauge hierarchy problem \cite{Randall:1999ee} (bulk fields in the 
Randall-Sundrum model have been considered in \cite{5DSM}). 
Through the AdS/CFT correspondence quantum loops in the bulk would 
correspond to $1/N_c$ corrections in the CFT side,
where $N_c$ counts the number of ``colors''.

Regarding again holography, we notice that the formalism of 
Refs. \cite{Parameswaran:2006db,Parameswaran:2007cb,Parameswaran:2009bt,articleA,articleB}  to compute the spectrum 
for codimension-two brane worlds that we reviewed here  can also be used to analyze spectral 
properties of holographic models in the  confined phase \cite{Witten:1998zw}: these are obtained from models
with one extra dimension (the holographic coordinate) by an additional compactified dimension and scalar, vector
and fermion fields in the  bulk can have a variety of uses ranging from condensed matter\footnote{See, however, \cite{Salvio} for
an alternative to compactification in this context.}
\cite{Nishioka:2009zj} to quantum chromodynamics \cite{Basu:2011yg}.

Finally, an interesting property of supersymmetry with two large extra dimensions is that
it could provide a link between the observed value of the cosmological constant and the scale at which modifications of 
gravity should occur, $1/r$ \cite{Callin:2005wi}. It would be interesting to know how gravity gets modified 
in the concrete supersymmetric model we discussed. In the absence of localized fluxes  graviton contributions
have been computed in \cite{Salvio:2009mp}, but the role of these fluxes, which are important for the dilaton
stabilization, remains an interesting target for future research.

\section*{Acknowledgements}
We would like to thank Cliff Burgess,  Leo van Nierop, Susha Parameswaran  and Matt Williams   
 for collaborations, 
 Hyun-Min Lee for much help 
 trying to diagonalize the supergravity sector in early stages of this work 
 and Riccardo Barbieri, Oriol Pujol\`as, Seifallah Randjbar-Daemi
and George Thompson for useful discussions. This work was partly 
supported by the EU ITN ``Unification in the LHC Era", contract PITN-GA-2009-237920 (UNILHC) 
and by MIUR under contract 2006022501.

\end{document}